\documentclass[%
 preprint,
 amsmath,amssymb,
 aps,
]{revtex4-1}

\usepackage{graphicx}
\usepackage{dcolumn}
\usepackage{bm}
\usepackage{mathtools}
\usepackage{breqn}
\usepackage{physics}
\usepackage{caption}
\usepackage{subcaption}
\usepackage{subcaption}
\usepackage{listings}
\usepackage[section]{placeins}
\usepackage{mathrsfs}

\begin{document}


\title{Infrared Generalized Uncertainty Principles Applied To LRS Bianchi I Quantum Cosmology}

\author{Daniel Berkowitz}
 \altaffiliation{Physics Department, Yale University.\\ daniel.berkowitz@yale.edu \\ This work is in memory of my parents, Susan Orchan Berkowitz, and Jonathan Mark Berkowitz}

\date{\today}

\begin{abstract}
We propose two higher order generalized uncertainty principles(GUPs) which predict a minimum uncertainty in momentum and apply the deformations that they entail of the Heisenberg algebra to one half of the phase space of the LRS Bianchi I models. After numerically solving the resultant Wheeler Dewitt equations we analyze our solutions and provide evidence that potential IR effects of quantum gravity could have played a role in selecting an overwhelmingly likely geometrical configuration that a quantum universe can possess. In addition we propose a GUP which predicts both a minimum uncertainty in momentum and a maximal measurable length scale which can be interpreted as a fixed maximum cosmological horizon. The results contained in this work provide further incentives to study what effects higher order GUP(s) have on quantum cosmology so we can obtain a better understanding of how quantum gravity could have impacted cosmological evolution. 
\end{abstract}

\pacs{Valid PACS appear here}
\maketitle


\section{\label{sec:level1}Introduction}
Many theories of quantum gravity propose\cite{amati1989can,gross1988string,gross1988string,hossenfelder2013minimal,kato1990particle} a minimal length scale which effectively acts as a natural UV cut off for field theories on space-time. This UV cut off results in a minimum length uncertainty $\Delta x$\cite{kempf1995hilbert,kempf1997minimal} of space-time which we can in principle measure and whose exact form can be obtained by introducing the following modified Heisenberg algebra
\begin{equation}\label{1}
[\hat{x}, \hat{p}]=i \hbar\left(1+B \hat{p}^{2}\right).
\end{equation}
The effects of this UV commutation  relation has been studied\cite{brau1999minimal,akhoury2003minimal,chung2019new,hamil2020new,vakili2007generalized,garattini2016cosmological,majumder2011dilaton} extensively in quantum mechanics and in quantum cosmology.

In this paper  though we will turn our attention to purely IR\cite{kempf1994quantum,kempf1997quantum} modifications of the Heisenberg algebra. Mirza and Zarei\cite{zarei2009minimal,chung2017ir}, and Chung, and Kim within the context of ordinary quantum mechanics studied the effects of the following first order modification to the Heisenberg algebra as we shall do for quantum cosmology. 
\begin{equation}\label{2}
[\hat{x}, \hat{p}]=i \hbar\left(1+a \hat{x}^{2}\right).
\end{equation}
Using this form of the commutation relation results in the following generalized uncertainty principle 
\begin{equation}\label{3}
\Delta x \Delta p \geq \frac{\hbar}{2}\left(1+a(\Delta x)^{2}+a\langle x\rangle^{2}\right),
\end{equation}
which we derived using 
\begin{equation}\label{4}
\Delta_{A}^{2} \Delta_{B}^{2} \geq\left(\frac{1}{2 i}\langle[\hat{A}, \hat{B}]\rangle\right)^{2}.
\end{equation}
By minimizing $\Delta p$ in (\ref{3}) we can compute the minimum uncertainty, or standard deviation that a sequence of momentum measurements on an ensemble of identical point particles could yield. The minimum uncertainty of momentum predicted by (\ref{3}) is $ \Delta p_{min}= \hbar\sqrt{a}$. This idea of introducing a minimum uncertainty in the momentum originates from the fact that for large distances when the curvature of space-time becomes important that no notion \cite{kempf1994quantum,kempf1997quantum} of a plane wave on a general curved space-time is known to exist; thus there appears to be a limit to the precision with which its momentum can be described.

As was done for the UV case \cite{pedram2012higher,nouicer2007quantum,chung2019new} we propose these two higher order IR modifications to the Heisenberg algebra which equals (\ref{2}) when expanded up to first order in $a$. The two higher order IR commutation relations we will be applying to quantum cosmology in our paper are the following,
\begin{equation}\label{5}
[\hat{x}, \hat{p}]=i \hbar e^{a \hat{x}^{2}},
\end{equation}
\begin{equation}\label{6}
[\hat{x}, \hat{p}]=i \hbar\sqrt{1+2 a  \hat{x}^{2}}.
\end{equation}
The modified commutation relation (\ref{5}) was influenced by Nouicer's higher order\cite{nouicer2007quantum} UV modification. 
As the reader can verify by following the methodology presented in the second section of \cite{chung2019new} these two relations respectively yield these generalized uncertainty principles,

\begin{equation}\label{7}
\Delta x \Delta p \geq \frac{\hbar}{2}\left(e^{a \left(\Delta x\right)^{2}}\right),
\end{equation}
\begin{equation}\label{8}
\Delta x \Delta p \geq \frac{\hbar}{2}\left(\sqrt{1+2 a \left(\Delta x\right)^{2}}\right).
\end{equation}
Using these GUPs we can obtain the minimum uncertainties in momentum that both (\ref{7}) and (\ref{8}) respectively predict. 

\begin{equation}\label{9}
\Delta p_{min}= \hbar\sqrt{\frac{a e^{1}}{2}},
\end{equation}
\begin{equation}\label{10}
\Delta p_{min}= \hbar\sqrt{\frac{a}{2}}.
\end{equation}
Because we are applying these GUP(s) to quantum cosmology we will understand x to be the scale factor $\alpha$ of the quantum Bianchi I models we will be studying. We will only have $\alpha$ obey our IR GUPs because it is a natural candidate to designate the spatial size of our system as $x$ is for one dimensional quantum mechanics. Another way of seeing this is by noticing that in the determinant $g$ in (\ref{14}) which gives a measure of our model's spatial volume that the anisotropic variable $\beta_{+}$ drops out.  

With that said we can go one step further and propose a commutation  relation which in addition predicts both a minimum uncertainty in momentum and predicts a maximum possible length scale which in terms of cosmology would correspond to a maximum measurable cosmic horizon. Following in the footsteps of \cite{pedram2012higher} we can write down 

\begin{equation}\label{11}
[\hat{\alpha}, \hat{p_{\alpha}}]=i \hbar \frac{1}{1-z\hat{\alpha}^{2}}.
\end{equation}
This relation predicts a minimum uncertainty $\Delta p_{\alpha \hspace{1mm} min}=\frac{3}{4} \sqrt{3} \sqrt{z} \hbar$ and a maximum scale factor of $\alpha_{max}=\frac{1}{\sqrt{z}}$. It would be of great interest to revisit the work of\cite{zarei2009minimal} and \cite{chung2017ir} using (\ref{11}), (\ref{6}) and (\ref{5}).

For our paper though we will apply (\ref{6}), (\ref{5}), and (\ref{2}) to the Bianchi I model we will introduce shortly. Once we describe our models we will present our numerical solutions to them and discuss how the different values of  $\Delta p_{\alpha \hspace{1mm} min}$ in (9) and (\ref{10}) could result in the differences we will observe in their wave functions. Afterwards we will provide some concluding remarks. 

\section{\label{sec:level2}The Model}

The LRS Bianchi I models possess the following metric when expressed in units where $c=\hbar=16 \pi G=1$

\begin{equation}\label{12}
d s^{2}=-N^{2}(t) d t^{2}+\alpha^{2}(t) e^{2 \beta_{i j}(t)} d x^{i} d x^{j}.
\end{equation}
In (\ref{12}) $\alpha(t)$ is the scale factor and ranges from $0 < \alpha(t) < \infty$ and $e^{2 \beta_{i j}(t)}$ is the following diagonal matrix  

\begin{equation}\label{13}
\beta_{i j}(t)=\operatorname{diag}(\beta(t)_{+}, \beta(t)_{+} ,-2 \beta(t)_{+}),
\end{equation}
where $\beta(t)_{+}$ is the Misner\cite{misner1969mixmaster,misner1969quantum} anisotropy variable. 

Proceeding to the Einstein-Hilbert action 

\begin{equation}\label{14}
\mathcal{S}=\int d^{4} x \sqrt{-g}(\mathcal{R}-\Lambda)
\end{equation}
where $g$ is the determinant, $\mathcal{R}$ is the Ricci scalar, and $\Lambda
$ is the cosmological constant results in the following Lagrangian in the minisuperspace variables $\left(\alpha,\beta_{+}\right)$

\begin{equation}\label{15}
\mathcal{L}= \frac{6 \alpha^{3} \dot{\beta_{+}^{2}}-6 \alpha \dot{\alpha}^{2}}{N}-\Lambda N\alpha^{3}.
\end{equation}
We choose the harmonic gauge for our lapse $N=\alpha^{3}$ and obtain the following Hamiltonian constraint 

\begin{equation}\label{16}
H=\frac{1}{24} \left(-\alpha^{2} p^{2}_{\alpha}+p^{2}_{\beta_{+}}\right)+\Lambda \alpha^{6}.
\end{equation}
Due to the fact that $\alpha$ is only defined on the positive portion of the real line it is not trivial to construct a self-adjoint operator out of it. This can be seen by finding imaginary eigenvalues of $-i \hbar\frac{\partial }{\partial \alpha}$ associated with eigenfunctions which can be normalized with respect to the standard inner product on the positive portion of the real line. As the reader can verify the aforementioned operator has $i$ as an eigenvalue and $e^{-\frac{\alpha}{\hbar}}$ as an eigenfunction, which is normalizable on the positive portion of the real line using the standard inner product. 

Using the suggestion\cite{bojowald2011quantum} given by Martin Bojowald we can remedy this problem by choosing a non-canonical pair of variables which satisfies the Poisson bracket 
\begin{equation}\label{17}
\poissonbracket{\alpha}{D}=\alpha
\end{equation}
where $D=\alpha p_{\alpha}$.
When quantized $\hat{D}$ only admits normalizable eigenfunctions with respect to $\int^{\infty}_{0}da$ when its eigenvalues are purely $\boldsymbol{real}$. Using these non canonical set of variables results in the following Hamiltonian. 
\begin{equation}\label{18}
H=\frac{1}{24} \left(-D^{2}+ p^{2}_{\beta_{+}}\right)+\Lambda \alpha^{6}.
\end{equation}
Quantizing (\ref{18}) results in (\ref{17}) turning into 
\begin{equation}\label{19}
[\hat{\alpha},\hat{D}]=i\hbar\hat{\alpha}
\end{equation}
where the $\alpha$ space representation of $\hat{D}$ is 
\begin{equation}\label{20}
\hat{D}=-\alpha i \hbar \frac{\partial}{\partial \alpha}.
\end{equation}
This method of quantization results in the following Wheeler Dewitt equation where we choose units such that $\hbar=1$ and implement the operator ordering $\hat{\alpha} \hat{p_{\alpha}} \hat{\alpha} \hat{ p_{\alpha}}$
\begin{equation}\label{21}
\alpha \frac{\partial}{\partial \alpha} \left(\alpha  \frac{\partial \Psi}{\partial \alpha}\right) - \frac{\partial^{2} \Psi}{\partial \beta^{2}_{+}} +24\Lambda \alpha^{6}\Psi=0.
\end{equation}
Deforming our Heisenberg algebra as we have done in (\ref{6}), (\ref{5}), and, (\ref{2}) results in the following three separable partial differential equations 
\begin{equation}\label{22}
\alpha\left(1+z\alpha^{2}\right) \frac{\partial}{\partial \alpha} \left(\alpha\left(1+z\alpha^{2}\right)  \frac{\partial \Psi_{1}}{\partial \alpha}\right) - \frac{\partial^{2}\Psi_{1}}{\partial \beta^{2}_{+}} +24\Lambda \alpha^{6}\Psi_{1}=0,
\end{equation}

\begin{equation}\label{23}
\alpha e^{z\alpha^{2}}\frac{\partial}{\partial \alpha} \left(\alpha e^{z\alpha^{2}} \frac{\partial \Psi_{2}}{\partial \alpha}\right) - \frac{\partial^{2} \Psi_{2}}{\partial \beta^{2}_{+}} +24\Lambda \alpha^{6}\Psi_{2}=0,
\end{equation}

\begin{equation}\label{24}
\alpha \sqrt{1+2z\alpha^{2}} \frac{\partial}{\partial \alpha} \left(\alpha \sqrt{1+2z\alpha^{2}} \frac{\partial\Psi_{3}}{\partial \alpha}\right) - \frac{\partial^{2} \Psi_{3}}{\partial \beta^{2}_{+}} +24\Lambda \alpha^{6}\Psi_{3}=0.
\end{equation}

These equations can be separated by requiring $\Psi_{i}$ to have the following forms 
\begin{equation}\label{25}
\Psi_{i}=e^{3ic\beta_{+}}f(\alpha)_{i}
\end{equation}
where c is an arbitrary constant. Utilizing (\ref{25}) results in a differential equation which can only be solved numerically whose solutions we will plot and analyze in the next section.

\onecolumngrid\
\begin{center}
\begin{figure}[h]
\centering
\begin{subfigure}{.4\textwidth}
  \centering
  \includegraphics[scale=.15 ]{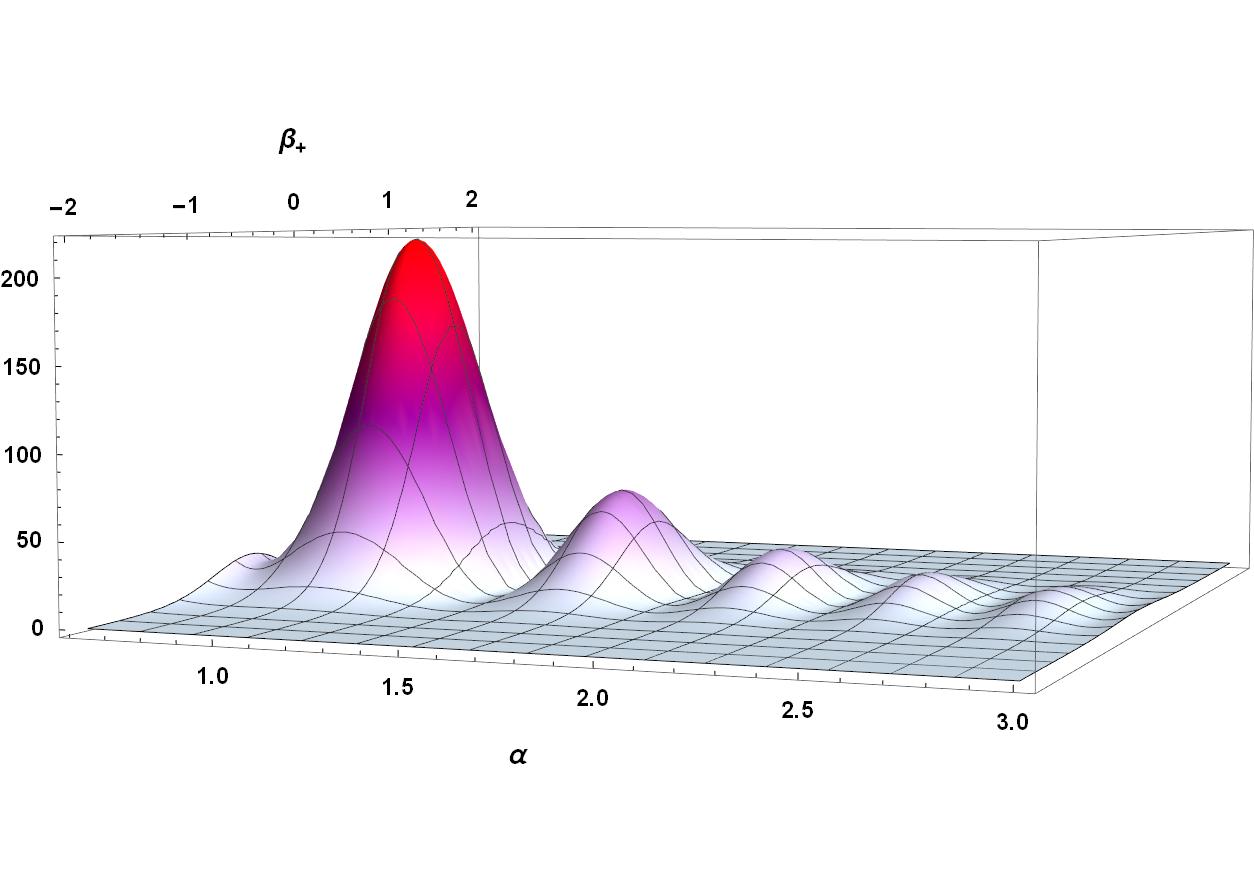}
  \caption{  $\abs{\Psi_{1}}^{2}$ \hspace{1mm} $ \Delta p_{\alpha \hspace{1mm} min}= \sqrt{z}$}
  \label{1a}
\end{subfigure}%
\begin{subfigure}{.4\textwidth}
  \centering
  \includegraphics[scale=.15 ]{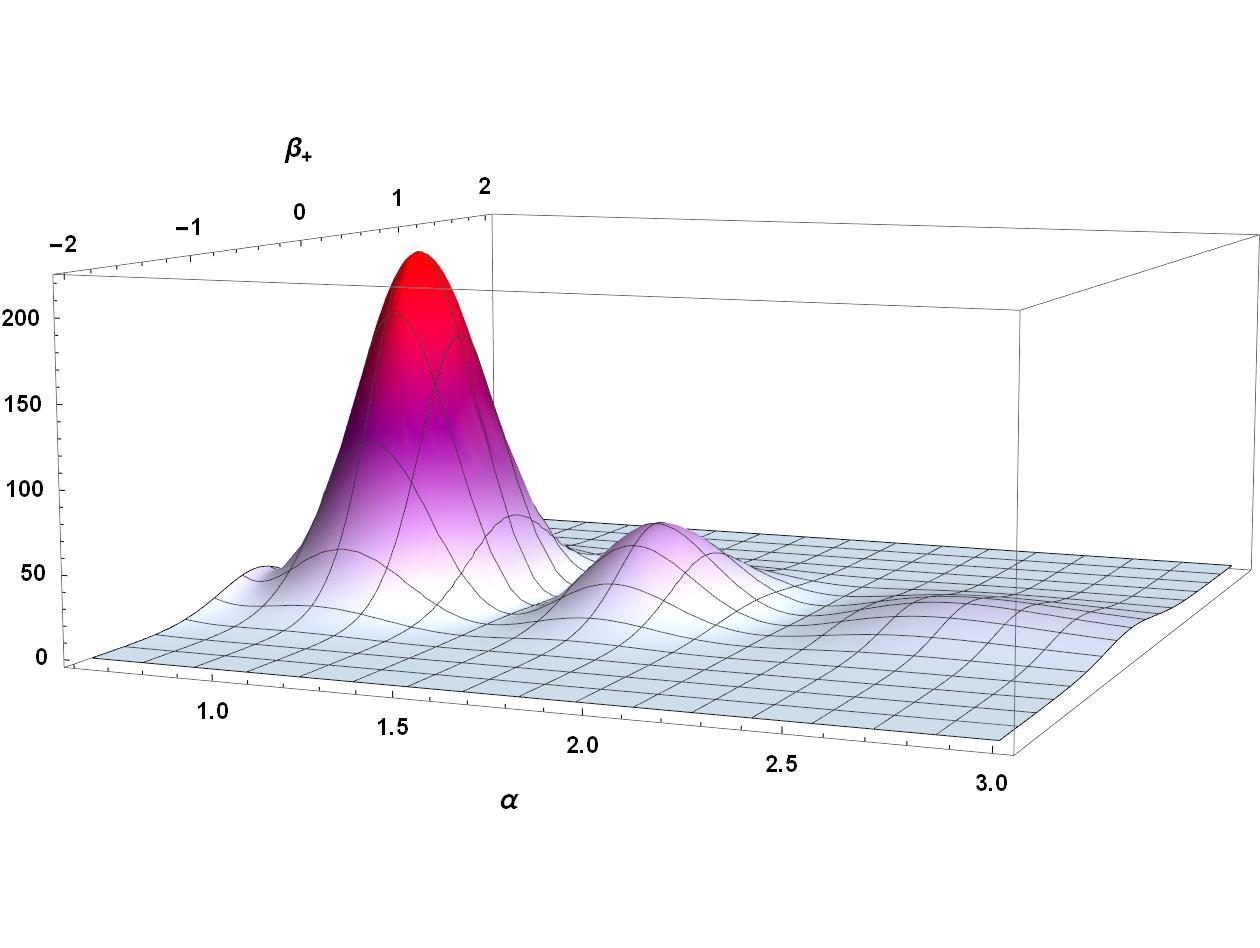}
  \caption{  $\abs{\Psi_{2}}^{2}$ \hspace{1mm} $ \Delta p_{\alpha \hspace{1mm} min}= \sqrt{\frac{e^{1}z}{2}}$ }
  \label{1b}
\end{subfigure}
\begin{subfigure}{.4\textwidth}
  \centering
 \ \includegraphics[scale=.15]{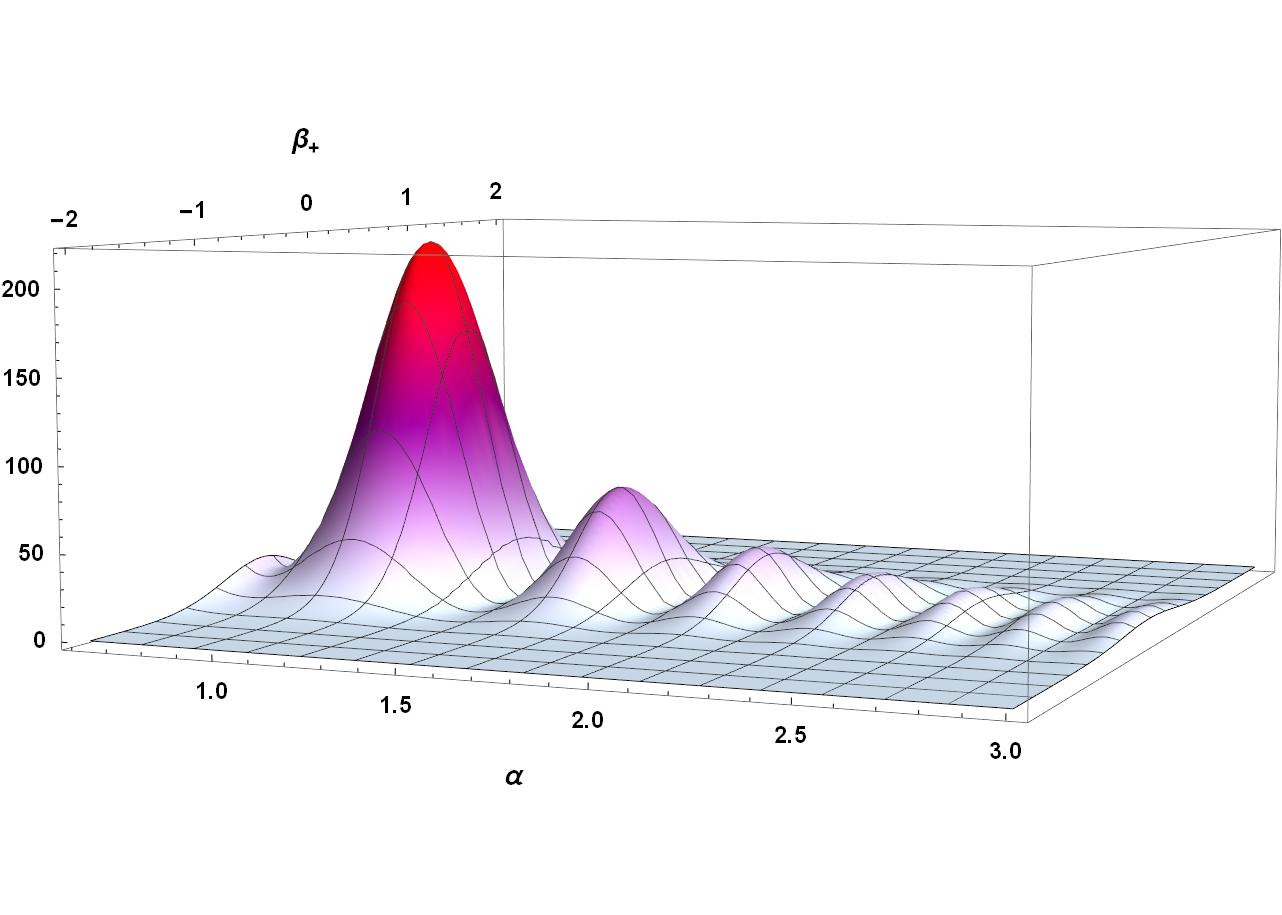}
  \caption{  $\abs{\Psi_{3}}^{2}$ \hspace{1mm} $ \Delta p_{\alpha \hspace{1mm} min}= \sqrt{\frac{z}{2}}$ }
  \label{1c}
\end{subfigure}%
\caption{These are plots for  $\abs{\Psi_{1}}^{2}$,  $\abs{\Psi_{2}}^{2}$, and  $\abs{\Psi_{3}}^{2}$ respectively when $z=.4$ and $\Lambda=1$ which we computed numerically using the initial conditions $f(1)=1$ and $f'(1)=-1$ where ' denotes the derivative with respect to $\alpha$. Figure 1(a) has 5 peaks present, figure 1(b) has 3 peaks present, and figure 1(c) has 6 peaks present. }
\label{fig1}
\end{figure}
\end{center}

\section{\label{sec:level3}Analysis Of Solutions To The IR Wheeler DeWitt Equations}
Before we discuss our solutions it should be said that typically GUP(s) are formulated on physical space. If the space in question possesses curvature then additional\cite{battisti2007big,scardigli1999generalized,scardigli2003generalized} steps are required to implement GUP(s) on it. However because the minisuperspace variables $\left(\alpha,\beta_{+}\right)$ fundamentally belong to a Lorentzian metric with vanishing Riemann curvature tensor as can be ascertained by the signs in (\ref{16}) the procedure we performed in the last section is valid. 

\begin{figure}[!ht]
\centering
\includegraphics[scale=.15]{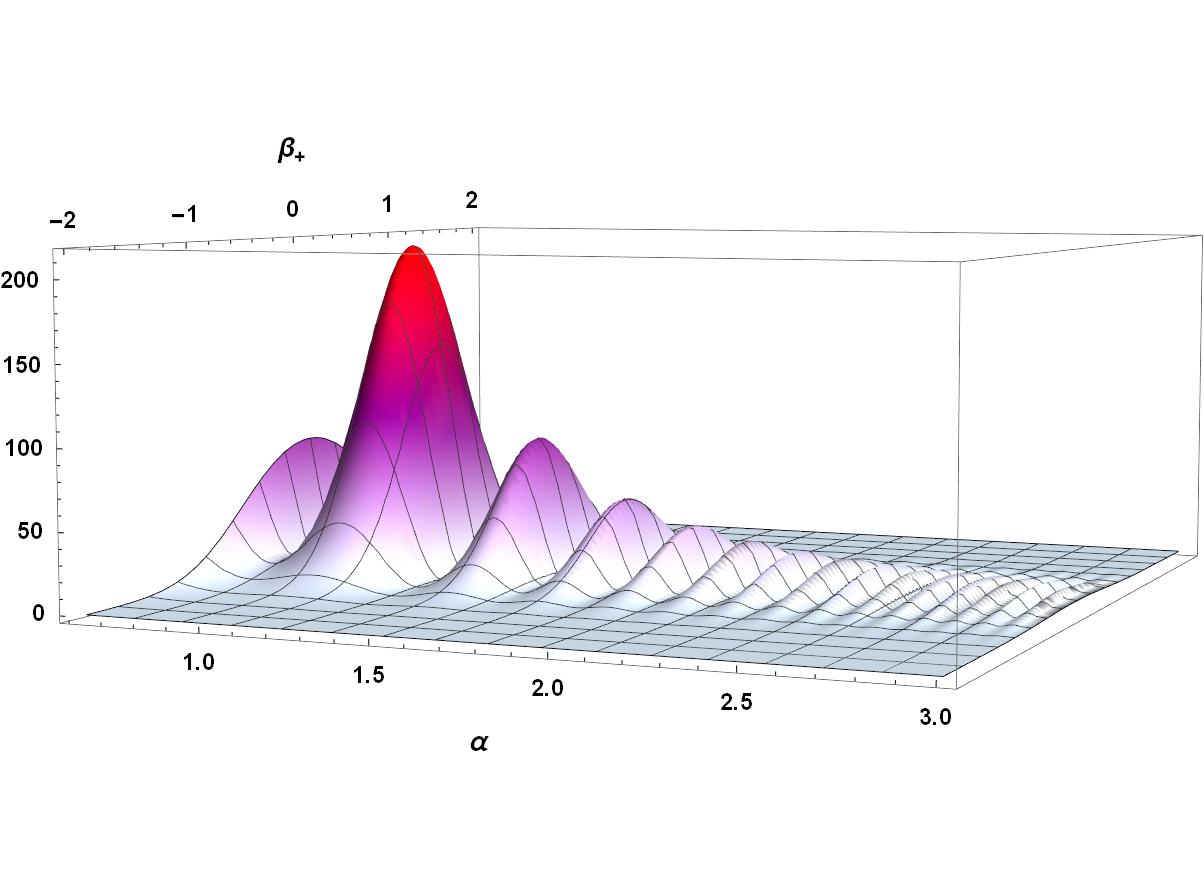}
\caption{ This is a plot for $\abs{\Psi_{i}}^{2}$ when $z=0$ and $\Lambda=1$ which we computed numerically using the initial conditions $f(1)=1$ and $f'(1)=-1$ where ' denotes the derivative with respect to $\alpha$.} 
\label{fig2}
\end{figure}

Using (\ref{25}) we will display four plots of the following wave functions 

\begin{equation}\label{26}
\Psi(\alpha,\beta_{+})_{i}=\int^{\infty}_{-\infty}e^{-1.5(c-1.1)^{2}}e^{3ic\beta_{+}}f_{i}(\alpha,c)dc
\end{equation}
where c is an integration constant that results from separating the variables. Figure (\ref{fig2}) will be when our IR cut off variable $z$ equals zero and thus results in the same wave functions for (\ref{24}), (\ref{23}), and (\ref{22}).  
The plots in figure (1) are for the three WDW equations (\ref{24}), (\ref{23}), and (\ref{22}) when $z=.4$. Via elementary quantitative analysis we expect IR effects to become important when our scale factor $\alpha$ approaches or surpasses $\frac{1}{\sqrt{z}}$.

In figure \ref{fig2} we see a wave function which possesses one central peak and multiple other smaller peaks. Each of these peaks represents a potential geometry given by a scale factor $\alpha$ and a measure of anisotropy $\beta_{+}$ that our quantum Bianchi I universe can possess. Because multiple peaks are present it is possible that our quantum Bianchi I universe can tunnel between different geometric configurations. This plot is similar to those that were constructed while studying the non-commutative Wheeler Dewitt equation\cite{garcia2002noncommutative,guzman2007non,aguero2007noncommutative} where the variable related to the scale factor ranged from $-\infty$ to $\infty$. 

We see in figure \ref{fig1} that the most prominent effect of increasing the minimum uncertainty of our momentum is that the lesser peaks which are present in figure\ref{fig1} are suppressed to varying degrees while the largest central peak remains unaffected. As can be seen the larger our minimum uncertainty in momentum is the more suppressed the non-central peaks are. This can be inferred by counting the number of non-central peaks present in our wave functions and noticing that the IR relations which predict larger minimum values of the uncertainty in the momentum suppress the non-central peaks more than those which predict smaller values. Thus for the wave functions we have plotted we can say that as the minimum uncertainty in momentum increases that the number of geometric states which our universes can tunnel in and out of decreases. To further explore what effects a minimum uncertainty in the momentum can induce onto the wave function of the universe(as discussed in \cite{hartle1983wave}) a greater variety of models will need to be studied and a more varied set of initial conditions will need to be considered.

\section{\label{sec:level4}Concluding Remarks}
In this paper we proposed two higher order GUPs which predict a minimum value of uncertainty in momentum and one other GUP which predicts both a minimum value of uncertainty in momentum and a maximal length scale. We then applied our two higher order generalized uncertainty principles and one first order GUP originally proposed by \cite{kempf1997quantum} to the quantum cosmology of the Bianchi I models. We found that IR aspects of quantum gravity can have prominent effects on cosmological evolution by choosing which geometric configuration is exceedingly likely that a universe will take on. 

An interesting avenue for future research would be to study quantum cosmology using both higher order IR and UV GUP which predict respectively a maximal length scale and a maximal momentum scale. A natural initial condition that such a wave function of the universe should possess is that $\Psi(\alpha_{max},\beta_{+})=0$. Doing so can further help us understand what effects quantum gravity could have played in the evolution of our universe. In addition the commutation  relations proposed here and in other works can be used to study classical cosmology as well by modifying the standard canonical $\poissonbracket{x}{p}$ Poisson brackets. The results and analysis present in this paper and others warrant the further study of how GUPs impact both quantum mechanical and cosmological systems. 

\section{\label{sec:level1}ACKNOWLEDGMENTS}
 
I am grateful to Professor Vincent Moncrief for valuable discussions at every stage of this work. I would also like to thank George Fleming for facilitating my ongoing research in quantum cosmology. Daniel Berkowitz acknowledges support from the United States Department of Energy through grant number DE-SC0019061. I also must thank my aforementioned parents.

\bibliography{IR.bib}

\end{document}